\documentclass{article}
\usepackage{amsmath,amssymb}

\begin{document}
\title{A universal amplitude ratio for the $q \leqslant 4$ Potts model
from a solvable lattice model.}
\author{Katherine A. Seaton
\thanks{
C. N. Yang Institute for
Theoretical Physics, State University of
New York, Stony Brook NY 11794-3840,USA}
\thanks{
Permanent address:
School of Mathematical and Statistical Sciences,
La Trobe University, Victoria 3086, Australia}
\thanks{email: k.seaton@latrobe.edu.au}}
\date{Novermber 15, 2001}

\maketitle

\begin{abstract} The universal amplitude ratio $R_{\xi}$ for
the $(q\leqslant 4)$-state Potts model in two dimensions is determined 
by using results for the dilute A
model in regime 1. The nature of the relationship between
the Potts model and the dilute A model, both related
to $\phi_{2,1}$ perturbed conformal field theory, is discussed.
\end{abstract}
%
%Keywords: {Potts model, dilute A model, universal amplitude ratio, percolation%}\\
%
%Running head: Universal amplitude ratio for Potts model
%\\

\noindent LTU Tech. report 2001-6\\
\noindent YITP-SB-01-69

%%%
\section{Introduction}
There has recently been interest in determining universal amplitude
ratios, familiar in statistical mechanics \cite{PHA}, using the 
techniques and results of perturbed conformal field theory.   
Since an integrable perturbation corresponds
to  the scaling limit of a two-dimensional lattice model
in statistical mechanics, these amplitudes
have found direct application to the Ising model \cite{delf},
the Potts model 
\cite{dc} and the tricritical Ising model \cite{FMS}, for example.
When the corresponding lattice model is solvable, or its universality
class contains a solvable counterpart, one would
hope to find some of the same amplitudes using the techniques and results of
the solvable model literature. Indeed, among the integrable field theory
results of reference \cite{delf} are
recovered the ``thermal'' amplitudes of the Ising model, known since
the seventies \cite{McWu}. In previous papers \cite{kas,SB7} universal 
amplitude ratios
for the  subleading magnetic and leading thermal
perturbations of the tricritical Ising model were calculated,
by considering their realization as members
of the dilute A$_L$ model hierarchy: the A$_3$ model in 
regime 1 and the A$_4$ model in regime 2, respectively. 
The amplitude ratios  obtainable
were confirmed to be identical to those found in references \cite{FMS}
and \cite{flzz}.

In this paper one universal amplitude ratio for the Potts
model, among those given in reference \cite{dc}, is determined 
from the dilute A model in regime 1 by utilizing a relationship
weaker than shared universality class. Preliminary results
for percolation ($q=1$) have been announced \cite{kas3}, and this
present paper completes the study.
Points of contact between the dilute A 
model and the $q\leqslant 4$ Potts model are outlined in Section \ref{models},
and expressions for the required thermodynamic quantities for 
the dilute A model are given.
In Section \ref{ratio} these connections and results are exploited to
determine an expression for the  
amplitude ratio $R_{\xi}^{+}$ of the Potts model for each integer value 
$1 \leqslant q
\leqslant 4$. For the Ising ($q=2$) case, it is also demonstrated that
this quantity can be determined from the Andrews-Baxter-Forrester \cite{ABF}
model.
The relationship of these results to  quantum field theory results
for the Potts model \cite{dc} 
is discussed in Section \ref{discuss}.

%%%% 
\section{The models}\label{models}
The dilute A$_L$ model \cite{WNS} is an $L$ state, 
interaction-round-a-face model 
which has been solved {\cite{WPSN} in four regimes,
two of which provide off-critical extensions of the unitary minimal conformal
field theories.  The model's adjacency diagram is that of A$_L$ with the
modification that a state may be adjacent to itelf on the lattice also. 
In regime 1 the
model is well-defined for integer $L \geqslant 2$.  Among others, one 
specification of regime 1
is the crossing parameter
\begin{equation}
\lambda=\frac{\pi L}{4(L+1)}.\label{cross}
\end{equation}
The central 
charge of the dilute A$_L$ model in this regime (or technically, at the
critical limit of the regime) is
\begin{equation}
c=1-\frac{6}{(L+1)(L+2)},\label{cA}
\end{equation}   
and the modular invariant partition function is $({\rm A}_L, {\rm A}_{L+1})$
\cite{R, obp} in the 
classification scheme of reference \cite{ciz}. In the scaling
limit the model realizes the perturbation $\phi_{2,1}$ of the minimal
unitary series $\mathcal{M}(L+1,L+2)$ \cite{WPSN}.
The elliptic nome $p$ which appears in the 
face weights of the
model corresponds to the coupling constant of the perturbation. For $L$
even the nome is thermal, and one should distinguish between regime 1$^{+}$
($p>0$) and regime 1$^{-}$ ($p<0)$.

The $q$-state Potts model for $q \leqslant 4$, on the basis of
its critical exponents determined by numerical and renormalization 
group studies
\cite{dn, nien}, has also been identified 
with the minimal
unitary series, by way of \cite{dot}
\begin{equation}
\sqrt{q}=2 \sin\frac{\pi(t-1)}{2(t+1)},\label{rule}
\end{equation}
when the central charge is written as
\begin{equation}
c=1-\frac{6}{t(t+1)}.\label{cpotts}
\end{equation}
The perturbation to which this model corresponds in the scaling limit
is again $\phi_{2,1}$. For both models this identification is made from 
the conformal weight in the leading term of the free energy 
(see (\ref{energy}) below).

That they are both identified with the same perturbation $\phi_{2,1}$ 
suggests a relationship between certain of 
the dilute A models and the Potts models
as given in Table \ref{relate}, or from inspection
of (\ref{cA}) and (\ref{cpotts}) by  naively setting $t=L+1$.
This is not to say that they are 
the same models, or even that they are in the same universality 
class.
In general the Potts models 
and the dilute A models have different internal
symmetries \cite{ro}, different numbers of ground states, and 
their order parameters
may be  associated with a different subset of the possible scaling fields 
of $\mathcal{M}(t, t+1)$.
This is the analogue of an idea discussed for three realizations
of the $\phi_{1,3}$
perturbation by Delfino \cite{delf2}. 
Just as there may be more than one $S$-matrix
in the field theory context associated with a particular perturbation,
here in the context of statistical mechanics there are two lattice models
with only some features in common.
For instance, the adjacency diagram of the 3-state Potts 
model has the symmetry of 
the D$_4$ Dynkin diagram, and the appropriate modular invariant partition 
function is $({\rm A}_4, {\rm D}_4)$
\cite{obp, ro}. The number of ground states of the dilute A$_L$ model grows
with $L$, but the 4-state Potts model (which we associate with
$L\to \infty$) has four ground states.
We should not, then,
expect universal quantities for the Potts
model which involve one-point functions or susceptibilities to
be obtainable via the dilute A model.
 
%
%Table goes here
\begin{table}[ht]
\caption{Potts models and dilute A$_L$ models which share a common
central charge and critical exponent $\alpha$.} \label{relate}
\begin{center}
\begin{tabular}{cccc}
\hline \hline
Central charge&Potts model&Dilute A model& $\alpha$\\\hline
$c \to 1$&$q=4$&$L \to \infty$&2/3\\
$c=4/5$&$q=3$&$L=4$&1/3\\
$c=1/2$&$q=2$&$L=2$&0\\
$c\to 0$&$q \to 1$&$L \to 1$&$-2/3$\\
\hline \hline
\end{tabular}
\end{center}
\end{table}
However, the universal amplitude ratio associated with the
specific heat and the correlation length is 
\cite{PHA}
\begin{equation}
R_{\xi}=A^{1/d} \xi_0 \label{Ramp}
\end{equation}
where $d$ is the dimension, 
$\xi_0$ is the leading term amplitude of the correlation length
\begin{equation*}
\xi \simeq \xi_0 \tau^{-\nu},
\end{equation*}
and  $A$ comes from the definition of
the amplitude of the specific heat
\begin{equation*}
C\simeq \frac{A}{\alpha} \tau^{-\alpha}.
\end{equation*}
Expressing $A$ in terms of the leading term coefficient  ${\mathcal
A}_f$ of the singular part of the free energy, 
\begin{equation*}
-f_{\rm s} \simeq {\mathcal
A}_f \tau^{2-\alpha},
\end{equation*}
it is possible to re-write (\ref{Ramp}) as:
\begin{equation}
R_{\xi}=\left[\alpha(1-\alpha)(2-\alpha) {\cal A}_f\right]^{1/d} \xi_0. 
\label{gen}
\end{equation}
The universality of $R_{\xi}$, i.e. its independence of metric factors 
associated
with the reduced temperature $\tau \propto T-T_{\rm c}$,  follows
from the scaling relation $2-\alpha =d \nu$. Of course, in what
follows for the lattice models we have $d=2$.

In the language of perturbed conformal field theory, the free energy and 
the correlation
length are related directly to the coupling constant $g$ of the perturbation 
and the
associated conformal weight $\Delta$:
\begin{equation}
f_{\rm s} \sim  g^{d/(d-2\Delta)} \qquad \qquad
\xi    \sim  g^{-1/(d-2\Delta)}\label{energy}.
\end{equation}
Thus when attention is confined to the amplitude ratio $R_{\xi}$,
the required quantities $\mathcal{A}_{f}$, $\xi_0$ and $\alpha$ (or 
equivalently $\Delta)$
relate solely to the perturbing operator. This operator is $\phi_{2,1}$
for both  dilute A in regime 1 and the Potts model, and any universal 
observable associated only
to it should be common \cite{delf2} for the points of contact between
the models,  as shown in 
Table \ref{relate}. 

The singular part of the free energy of the dilute A$_L$ 
model in regime 1 has been determined using the inversion 
relation \cite{WPSN} and exact perturbative \cite{BS,BS2}
approaches.  The leading term is \cite{WPSN}:
\begin{equation*}
f_{\rm s}\sim \left\{\begin{array}{lll}
p^2 \ln(p)&&L=2\\  
p^{4(L+1)/3L}&&L\geqslant 3 \end{array} \right. ,
\end{equation*}
 so that for $L$ even 
\begin{equation}
\alpha = \frac{2(L-2)}{3L}. \label{alf}
\end{equation}
 Apart from when $L=2$, the coefficient \cite{kas3} is:
\begin{equation}
\mathcal{A}_f =\frac{4 \sqrt{3} \sin(2\pi(L-1)/3L)}{\sin(\pi(L-2)/3L)}.
\label{free}
\end{equation}

The leading term of the correlation length \cite{BS, BS2} is
\begin{equation}
\xi^{-1}\simeq 4 \sqrt{3}\  p^{2(L+1)/3L}. \label{lead}
\end{equation}
Strictly speaking this latter expression was determined for $L$ odd, where it 
applies both when $p>0$ and $p<0$, but there is good reason to believe that it also
applies to the high temperature regime for $L$ even. The amplitude ratio
found in this paper is thus $R^{+}_{\xi}$, that is, it applies coming from
above the critical temperature. 

Substituting the results (\ref{alf})-(\ref{lead}) into (\ref{gen}), 
the general expression for
this particular universal amplitude ratio of the dilute A$_L$ models in
regime 1 is:
\begin{equation}
R^{+}_{\xi}= \left[\frac{2(L-2)(L+1)(L+4)}{27 \sqrt{3}L^3}
\frac{\sin(2\pi(L-1)/3L)}{\sin(\pi(L-2)/3L)}
\right]^{\frac{1}{2}}.
\label{uni}
\end{equation}
Though the discussion above focussed on thermal fields,
this expression represents a universal quantity for all $L$;
for $L$ odd the nome is  magnetic-field-like.

Since $L$, like $q$, labels the number of states in the model, 
it would seem it should always be an integer at least 
equal to 2. However, it has long been realized that $q$ can
be treated as a continuous variable, when
the Potts model is formulated in terms of the random
cluster model {\cite{FK}. In particular, by taking the 
limit $q \to 1$ results for percolation can be obtained.
In a similar way we will take $L\to 1$ 
in the dilute A model, as foreshadowed in
Table \ref{relate}; in the expression for $R_{\xi}^{+}$ 
(\ref{uni}) there is no impediment to 
letting $L$ run through
all natural numbers. Alternatively
one can think of the crossing parameter $\lambda$ given in (\ref{cross})
varying quasi-continuously from $\pi/8$ to $\pi/4$. Technical details
to do with treating $L$ in this way will be mentioned as necessary, as the 
four values relevant to the Potts model are now
considered.

\section{The universal amplitude ratio}\label{ratio}
\subsection{Potts model with $q = 3, 4$}
It is now straight-forward to determine the amplitude
ratio between the specific heat, or singular part of the free energy, 
and the correlation
length for the $q=3$ state Potts model. Setting $L=4$ in (\ref{uni}):
\begin{equation}
R_{\xi}^{+}=\left[ \frac{5}{27 \sqrt{3}}  \right]^{\frac{1}{2}}.\label{three}
\end{equation}

To determine the corresponding amplitude ratio for the $q=4$ state
Potts model, the limit $L\to \infty$ is taken in (\ref{uni}).
The result obtained is
\begin{equation}
R^{+}_{\xi}=\left[\frac{2}{27 \sqrt{3}}\right]^{\frac{1}{2}}.\label{four}
\end{equation}

\subsection{The Ising model in zero magnetic field, or $q=2$}
It is hardly necessary  to obtain an expression 
for $R_{\xi}^{+}$ for the 2-state Potts model via the dilute A model, since
the result is exactly known from the equivalence of this case to the (thermal)
Ising model. The results obtained by field theoretic approaches 
\cite{delf,dc} have already been shown to agree with the 
lattice Ising model values \cite{McWu}. In the interests of completeness, then,
let us confirm that the known value 
\begin{equation*}
R^{+}_{\xi}=\frac{1}{\sqrt{2 \pi}}
\end{equation*}
for the Ising model in zero magnetic field is recovered from the
dilute A$_2$ model in regime 1.

The expression (\ref{free}) for the coefficient $\mathcal{A}_f$
of the dilute A$_L$ model does not apply when $L=2$; in this case
correct treatment of the expression for the partition function in
\cite{WPSN} or \cite{BS} gives 
\begin{equation*}
f_{\rm s} \simeq \frac{12}{\pi} p^2 \ln( p).
\end{equation*}
Modifying
the definition of the amplitude $C \simeq A \ln (p)$ as is appropriate
for the logarithmic divergence, one obtains as expected
\begin{equation*}
R_{\xi}^{+}=\left[ 2 \mathcal{A}_f\right]^{\frac{1}{2}} \xi^{+}_{0} =
\frac{1}{4\sqrt{3}} \left[\frac{24}{\pi}\right]^{\frac{1}{2}}=
\frac{1}{\sqrt{2\pi}}.
\end{equation*}

However, the general dilute A$_L$ expression (\ref{uni}) for $R_{\xi}^{+}$ is 
well-behaved at $L=2$. Taking the limit $L\to 2$ gives, correctly,
\begin{equation}
R^{+}_{\xi}= \left[\frac{2(L+1)(L+4)}{9 \pi \sqrt{3}L^2}
\cos(\pi(L-4)/6L)
\right]^{\frac{1}{2}}_{L=2}=\frac{1}{\sqrt{2\pi}}.\label{two}
\end{equation}

Incidentally, for this Ising model case, which is related to
the minimal unitary conformal field theory $\mathcal{M}(3,4)$, 
the scaling field 
$\phi_{2,1}=\phi_{1,3}$, which can be seen from the identity for conformal
weights 
\begin{equation*}
\Delta_{j,k}^{(4)}=\Delta^{(4)}_{3-j, 4-k}.
\end{equation*}
The $(r-1)$-state models of 
Andrews, Baxter
and Forrester \cite{ABF} are known \cite{hu} to realise the $\phi_{1,3}$
perturbation of the
minimal unitary series $\mathcal{M}(r-1,r)$ and for $r=4$
should also
give the 2-state Potts amplitude ratio under consideration.
 
The free energy and correlation length of the ABF models,
obtained \cite{ABF} from the 8-vertex model results \cite{Baxter}, are:
\begin{align}
f_{\rm s} &\simeq -4\cot(\pi^2/2\lambda) \tau^{\pi/2\lambda}\label{freeabf}\\
\xi^{-1}&\simeq 4 \tau^{\pi/4\lambda}. \label{xiabf}
\end{align}
However, the crossing parameter is $\lambda={\pi}/{r}$ for the ABF models,
and for $r$ even the free energy (\ref{freeabf}) should properly 
be modified with a logarithmic factor,
and the coefficient re-calculated. Instead, simply 
constructing the amplitude ratio (\ref{gen})
of the coefficients in (\ref{freeabf}) and (\ref{xiabf}) and taking $r \to 4$
by the approach  used to obtain (\ref{two}) 
for the dilute A model:
\begin{align*}
R_{\xi}^{+}&=\lim_{r\to 4} \left[
\frac{(r-4)(r-2)r}{32}\ \frac{\cos(\pi r/2)}{\sin(\pi r/2)}
\right]^{\frac{1}{2}}\\
&=\frac{1}{2}\lim_{r\to 4} \left[
\frac{(r-4)}{\sin(\tfrac{\pi r}{2}-2 \pi)}
\right]^{\frac{1}{2}}=\frac{1}{\sqrt{2 \pi}}.
\end{align*}

\subsection{Percolation, or $q \to 1$}
The percolation result, though previously presented \cite{kas3},
is reiterated here for completeness.
A review of the relationship between the $q$-state Potts model
and percolation from the point of view of universal
amplitude ratios is given in reference \cite{dc}. 
The appropriate object of interest for percolation is the ratio
\begin{equation*}
\tilde{R}^{+}_{\xi}=\lim_{q\to 1}\frac{R^{+}_{\xi}}{(q-1)^{1/2}}.
\end{equation*}

To obtain $\tilde{R}^{+}_{\xi}$  
from the dilute A model, we put $t=L+1$ in the 
expression (\ref{rule}) for $q$, 
and then
apply trigonometric identities to $(q-1)$:
\begin{equation*}
q-1=4 \sin \left(\frac{\pi (2L+1)}{3(L+2)} \right)
\sin \left(\frac{\pi (L-1)}{3(L+2)}\right).
\end{equation*}
Although there is a factor in  the numerator of (\ref{uni}) which
becomes zero at $L=1$,
we see that its ratio with $(q-1)$ will be finite in the limit $L\to 1$,
so that
\begin{equation}
\tilde{R}^{+}_{\xi}=\left[\frac{(L-2)(L+1)(L+4)(L+2)}{27 \sqrt{3}L^4
\sin\left(\frac{\pi (2L+1)}{3(L+2)}\right)\sin\left(\frac{\pi(L-2)}{3L}
\right)}
\right]^{\frac{1}{2}}_{L=1}=\left[\frac{40}{27\sqrt{3}}\right]^{\frac{1}{2}}.
\label{one}
\end{equation}

%%%
\section{Discussion}\label{discuss}

In 1984 Kaufman and Andelman \cite{KA} presented an
argument that the specific
heat amplitude ratio (above and below $T_{\rm c}$) is
$A_{+}/A_{-}=1$ for the $q$-state Potts model ($q\leqslant 4$). 
The free energy expression (\ref{free})
applies for both signs of $p$, so that this value of the
universal amplitude ratio for the specific heat holds for the
dilute A model for all $L$, including the special cases applicable
to the Potts model.

Moreover, an expression was proposed in reference \cite{KA}
for the $q$-dependence of the
amplitude of the singular part of the free energy of the
Potts model for $q \leqslant 4$, which accounted for its
known divergences and zeroes and which we will denote $A_{\rm KA}$. 
Subsitituting
$t=L+1$ in (\ref{rule}) and this 
then into $A_{\rm KA}$ it can be shown on
rearranging and comparison with (\ref{free}) that
\begin{equation*}
A_{\rm KA}=\frac{b(q)}{6} \mathcal{A}_{f}.
\end{equation*}
Here $b(q)$ is a positive, slowly-varying function allowed for in \cite{KA}
so that $A_{\rm KA}$ and $\mathcal{A}_f$ must 
have common zeroes and divergences.
We have already observed, in constructing the amplitude
ratio, that $\mathcal{A}_{f}$  is
divergent at $q=2$ and zero at $q=1$.

It was remarked below expression (\ref{uni}) that it  represented a
universal quantity for the dilute A$_L$ model for all $L$.
It is related in a straight-forward manner to the universal quantity
considered in quantum field theory $\varepsilon m^{-2}$, where $\varepsilon$
is the bulk vacuum energy and the mass $m$ of the field
theory is the inverse 
of the correlation length $\xi$ in the scaling limit of the lattice model. The
quantity (see (\ref{free}) and (\ref{lead}))
\begin{equation}
-f_{\rm s} \xi^{2}=\frac{\sin(2\pi(L-1)/3L)}{4\sqrt{3}\sin(\pi(L-2)/3L)}
\label{uni2}
\end{equation}
agrees exactly (when the various notations are
translated) with $\varepsilon m^{-2}$ calculated for the $\phi_{2,1}$ perturbed
theory by the thermodynamic Bethe ansatz \cite{F} and two-kink form factor 
approach \cite{dc} based on the $S$-matrix of Chim
and Zamolodchikov \cite{chim}.

Thus the algebraic expressions for $R_{\xi}^{+}$ for the Potts
model
calculated in this paper from the dilute A model agree precisely
with those obtained implicitly by Delfino and Cardy \cite{dc}.
The numerical
values given in Tables 3 and 5  of reference \cite{dc} 
(for comparison with previous numerical results for the
lattice Potts model) use the `second moment' correlation
length, which differs by a few percent from the `true'
correlation length used here. This can be seen in Table \ref{numbers}
where the second moment values \cite{dc} for $R_{\xi}$ are reproduced
together with evaluations of the exact expressions 
(\ref{three}), (\ref{four}), (\ref{two}) and (\ref{one}).

%

%table goes here
\begin{table}[ht]
\caption{The Potts model universal
amplitude ratio $R_{\xi}^{+}$, determined from quantum field theory in the
two-kink approximation to the form-factors, using the second
moment correlation length (by Delfino and Cardy), 
and from special cases of the dilute A model using the
true correlation length (this paper).} \label{numbers}
\begin{center}
\begin{tabular}{ccc}
\hline \hline
Potts model&Two-kink approx.&Dilute A model\\
\hline
$q=4$       & 0.2052  & $2^{1/2}3^{-7/4}=0.20680\ldots$\\
$q=3$ & 0.3262 & $5^{1/2}3^{-7/4}=0.32698\ldots$\\
$q=2$ & 0.3989 & $2^{-1/2}\pi^{-1/2}=0.39894\ldots$\\
$q=1$&0.926\phantom{0}&$2^{3/2}5^{1/2}3^{-7/4}=0.92484\ldots$\\
\hline \hline
\end{tabular}
\end{center}
\end{table}
 The authors of reference \cite{dc} have further
obtained numerical values for other universal amplitude
ratios for the Potts models which do not appear to be accessible from
solvable lattice models.  Their various results are new for $q=3$, $4$ and 
improve results for percolation ($q \to 1$) from Monte-Carlo or 
series enumeration techniques for the lattice Potts model itself.
The good accuracy of the field theoretic approach was previously discussed
in the context of self avoiding walks \cite{cm}. 
Nevertheless, it is hoped 
that this calculation in the solvable model context, though limited to one
of the amplitudes, is of interest to field theorists and statistical 
mechanists alike.

\section*{ACKNOWLEDGMENTS}
This work was completed during sabbatical (OSP) leave 
spent at the Department of Mathematics and Statistics, University
of Melbourne and at SUNY Stony Brook. The author thanks Barry McCoy
for his encouragement and hospitality, and acknowledges helpful 
correspondence from Aldo Delfino.

%%%%%%%

\end{document}